\documentclass[10pt,journal,twocolumn]{IEEEtran}
\usepackage{cite}
\usepackage{amsmath,amssymb,amsfonts}
\usepackage{graphicx}
\usepackage{textcomp}
\usepackage{amsthm}
\usepackage{optidef}

\newtheorem{theorem}{Theorem}[section]
\newtheorem{lemma}[theorem]{Lemma}
\newtheorem{remark}{Remark}

\usepackage{multicol, blindtext}
\usepackage{lipsum}
\usepackage{afterpage,lipsum,capt-of,graphicx}
\usepackage{caption}
\usepackage{cuted}
\usepackage{algorithmicx}
\usepackage{algpseudocode}
\usepackage{algorithm}
\algblockdefx{MRepeat}{EndRepeat}{\textbf{repeat}}{}
\algnotext{EndRepeat}
\enlargethispage{-2cm}
\usepackage{mathtools}
\usepackage{pbox}
\usepackage{mathtools}
\usepackage{subcaption}
\usepackage{refcount}
\usepackage{flushend}

\usepackage{relsize}

\begin{document}

\title{ \Large \bf Distributed Linear Quadratic Gaussian for Multi-Robot Coordination with Localization Uncertainty}

\author{Tohid Kargar Tasooji, Sakineh Khodadadi

\thanks{ Tohid Kargar Tasooji is with the Department of Aerospace Engineering,
Toronto Metropolitan University, Toronto, ON M5B 2K3, Canada (e-mail:
tohid.kargartasooji@torontomu.ca). Sakineh Khodadadi is with the Department of Electrical and Computer
Engineering, University of Alberta, Edmonton,AB,  AB T6G 1H9, Canada (email: sakineh@ualberta.ca).}
}

\maketitle

\begin{abstract} \small\baselineskip=9pt
This paper addresses the problem of distributed coordination control for multi-robot systems (MRSs) in the presence of localization uncertainty using a Linear Quadratic Gaussian (LQG) approach. We introduce a stochastic LQG control strategy that ensures the coordination of mobile robots while optimizing a performance criterion. The proposed control framework accounts for the inherent uncertainty in localization measurements, enabling robust decision-making and coordination. We analyze the stability of the system under the proposed control protocol, deriving conditions for the convergence of the multi-robot network. The effectiveness of the proposed approach is demonstrated through experimental validation using Robotrium simulation-experiments, showcasing the practical applicability of the control strategy in real-world scenarios with localization uncertainty.
\end{abstract}

\section{Introduction} \IEEEPARstart{I}{N} recent years, the coordination and control of multi-robot systems (MRSs) have become critical research areas, driven by their diverse applications in autonomous systems. These include traffic control and management, search and rescue operations, collaborative exploration, and formation control for autonomous vehicles \cite{1, 2, 3, 4, 28, 29, 30, 31, 32, 33, 39, 40, 41, 42}. A central task in such applications is achieving rendezvous, where multiple robots converge to a common location or state in a distributed fashion. Rendezvous is a critical problem in multi-agent systems (MASs), particularly in scenarios where cooperation among agents is required without centralized control, and where the environment is dynamic and uncertain.

In the context of multi-robot rendezvous, much of the existing research focuses on consensus control, where agents aim to agree on a common state, such as position or velocity, often with idealized assumptions of perfect sensor measurements and communication channels. While these works have laid the groundwork for practical multi-robot coordination, they do not fully address the challenges posed by real-world uncertainties, such as noisy sensor measurements, imperfect localization, and communication delays, which are commonly encountered in mobile robot applications.

Localization is one of the key factors that affects the effectiveness of rendezvous in multi-robot systems. In practice, each robot may rely on sensors such as cameras, IMUs (inertial measurement units), and encoders to estimate its position, but these measurements are often corrupted by noise and errors. Without compensating for these uncertainties, achieving reliable rendezvous becomes challenging. Furthermore, errors in the localization process can lead to discrepancies between the true position of the robots and their estimated positions, hindering their ability to converge to a common location accurately.

To address these challenges, this paper introduces a novel approach based on the Linear Quadratic Gaussian (LQG) control framework for multi-robot rendezvous. The LQG controller is specifically designed to account for the uncertainties introduced by noisy localization data. We propose a distributed approach to LQG control, which enables each robot to independently calculate its control inputs while taking into account the noisy state estimates obtained from sensor fusion techniques such as Kalman filtering. This approach ensures that the multi-robot system can achieve rendezvous despite the presence of noise in sensor data, which is a common issue in real-world applications.

The proposed LQG-based rendezvous control design uses a quadratic cost function, which balances the trade-off between achieving the rendezvous and minimizing energy consumption, ensuring optimal performance under practical constraints. The optimal control gains are derived using the Bellman equation, which allows us to address both the performance and robustness requirements of multi-robot rendezvous.

The contributions of this paper are as follows: \begin{enumerate} \item We introduce a distributed LQG control framework for multi-robot rendezvous that explicitly incorporates localization uncertainty. This approach allows the robots to achieve rendezvous even with noisy sensor measurements, which are prevalent in practical mobile robot applications. Unlike traditional consensus protocols that assume accurate state measurements, our method takes into account the noisy nature of sensor data and adjusts the control law accordingly. \item We provide a rigorous stability analysis of the proposed rendezvous control protocol. We analyze the effect of localization error and sensor noise on the upper bound of the rendezvous error and derive conditions under which the robots will successfully converge to a common position. This analysis contributes to the theoretical understanding of rendezvous in stochastic environments. \item We validate the effectiveness of the proposed control strategy through Robotrium simulation-experiments, demonstrating that the proposed method not only achieves rendezvous but also optimizes the performance index, even under realistic constraints such as limited energy resources, communication delays, and bandwidth restrictions. These experiments serve as a bridge between the theoretical analysis and practical implementation, providing evidence of the robustness and reliability of the proposed solution in real-world scenarios. \end{enumerate}

This paper aims to contribute to the ongoing research in multi-robot systems by addressing the problem of rendezvous in the presence of sensor noise and localization uncertainty. The proposed distributed LQG control framework offers a promising solution for achieving reliable and optimal rendezvous in stochastic environments, which is crucial for the deployment of autonomous mobile robots in real-world tasks.

The rest of this paper is organized as follows. Section II provides the problem formulation and background. Section III presents the design of the stochastic LQG controller. In Section IV, we analyze the stability of the multi-robot system under the proposed controller. Section V presents the experimental results from Robotrium simulations to validate the theoretical findings. Finally, Section VI concludes the paper and discusses future research directions.

\section{Preliminaries}

\subsection{Notation}
Throughout this paper, we denote by $\mathbb{R}^n$ the set of $n$-dimensional real vectors and by $\mathbb{R}^{p \times q}$ the set of real matrices with $p$ rows and $q$ columns. For a matrix $A$, $\|A\|$ represents its induced matrix 2-norm, and $A^T$ denotes its transpose. The identity matrix of an appropriate size is denoted by $I$. For matrices $A \in \mathbb{R}^{m \times n}$ and $B \in \mathbb{R}^{p \times q}$, the Kronecker product $A \otimes B$ is defined as:
\begin{equation*}
A \otimes B =
\begin{bmatrix}
a_{11}B & \cdots & a_{1n}B \\
\vdots & \ddots & \vdots \\
a_{m1}B & \cdots & a_{mn}B
\end{bmatrix}.
\end{equation*}
We write $A > 0$ (or $A \ge 0$) to indicate that $A$ is positive definite (or positive semidefinite), meaning all its eigenvalues are positive (or nonnegative).

To model inter-robot communication, we represent the network of $N$ robots as a directed graph $\mathcal{G} = (\mathcal{V}, \mathcal{E}, \mathcal{A})$, where $\mathcal{V} = \{1, 2, \dots, N\}$ is the set of robot indices. A directed edge $(i,j) \in \mathcal{E}$ indicates that robot $i$ can transmit information to robot $j$. The adjacency matrix $\mathcal{A} = \lfloor a_{ij} \rfloor$ is defined by:
\[
a_{ij} = \begin{cases}
1, & \text{if } (i,j) \in \mathcal{E}, \\
0, & \text{otherwise},
\end{cases}
\]
with $a_{ii} = 0$ for all $i$. The Laplacian matrix $\mathcal{L} = [l_{ij}]$ of the graph is given by
\[
l_{ii} = \sum_{j \neq i} a_{ij}, \quad l_{ij} = -a_{ij} \quad (i \neq j),
\]
which captures the structure of inter-robot interactions.

\subsection{Problem Formulation}
Consider a team of $N$ mobile robots operating in a two-dimensional environment. Each robot is characterized by its position $(x,y)$ and heading $\theta$, where $\theta$ is measured in degrees relative to the positive $x$-axis (within the interval $[-180^\circ, 180^\circ]$). Assuming that the motion along the $x$ and $y$ axes is decoupled, the kinematic model for the $i$th robot is given by:
\begin{equation} \label{eq:dynamics}
\left\{
\begin{aligned}
\dot{x}_{i}(t) &= u^{i}_{x}(t) = v_{i}(t)\cos\big(\theta_{i}(t)\big), \\
\dot{y}_{i}(t) &= u^{i}_{y}(t) = v_{i}(t)\sin\big(\theta_{i}(t)\big),
\end{aligned}
\right.
\quad i = 1, \dots, N,
\end{equation}
where $x_{i}(t)$ and $y_{i}(t)$ denote the position components of the $i$th robot, and $u^{i}_{x}(t)$ and $u^{i}_{y}(t)$ are the control inputs along the $x$ and $y$ axes, respectively. The scalar $v_{i}(t)$ represents the linear velocity of the robot.

 In this work, each robot samples its state and broadcasts noisy measurements to its neighbors. Due to the presence of sensor noise and localization uncertainties, onboard sensor fusion techniques—such as the Extended Kalman Filter (EKF)—are utilized to generate improved state estimates. The EKF fuses measurements from sensors like odometry and inertial measurement units (IMU), thereby enhancing the accuracy of the robot’s pose estimation.

The primary objective of this work is to design a distributed Linear Quadratic Gaussian (LQG) controller that drives the multi-robot system to achieve rendezvous, i.e., convergence of all robots to a common position, while optimizing a quadratic performance index. Mathematically, the desired rendezvous objective is expressed as:
\[
\lim_{t \to \infty} \|\hat{x}_{i}(t) - \hat{x}_{j}(t)\| = 0 \quad \text{and} \quad \lim_{t \to \infty} \|\hat{y}_{i}(t) - \hat{y}_{j}(t)\| = 0, \] \[  \quad \forall \, i,j \in \{1,\dots,N\},
\]
where $\hat{x}_{i}(t)$ and $\hat{y}_{i}(t)$ are the EKF-based state estimates for the $i$th robot. The proposed distributed LQG controller not only ensures robust rendezvous in the presence of localization uncertainty but also minimizes energy consumption and other performance costs, thereby meeting practical deployment requirements in multi-robot applications.

\begin{remark}
In this study, we consider a synchronous and periodic sampling framework for the multi-robot system. Let $\{t_{k}\}_{k=0}^{\infty}$ denote the discrete time instants at which each robot samples its state and transmits the information to its neighbors, with a constant sampling interval $h$, i.e., $t_{k+1}-t_{k}=h \geq 0$. In our experimental implementation, communication among robots is facilitated via a Bluetooth network, which limits the maximum sensor sampling rate to 10 Hz. This periodic sampling assumption simplifies the analysis and implementation of the distributed LQG rendezvous controller.
\end{remark}
Discretizing the continuous-time dynamics in (1) using a zero-order hold and a periodic sampling interval $h$, we obtain the following discrete-time model for the $i$th robot:
\begin{equation} \label{eq:discreteDynamics}
\left\{
\begin{aligned}
x_i(k+1) &= x_i(k) + h\, u_{x_i}(k) + w_{x_i}(k), \\
y_i(k+1) &= y_i(k) + h\, u_{y_i}(k) + w_{y_i}(k),
\end{aligned}
\right.
\quad i=1,\ldots,N,
\end{equation}
where $x_i(k)$ and $y_i(k)$ denote the robot's position components at the $k$th sampling instant, $u_{x_i}(k)$ and $u_{y_i}(k)$ are the control inputs along the $x$ and $y$ axes, and $w_{x_i}(k)$ and $w_{y_i}(k)$ represent independent zero-mean Gaussian process noise terms.

To address the effects of sensor noise and localization uncertainty, we adopt an Extended Kalman Filter (EKF) for state estimation. The discrete-time EKF update equations for the $i$th robot along the $x$-axis are given by:
\begin{equation} \label{eq:ekfX}
\begin{aligned}
\hat{x}^i_{k|k} &= A\,\hat{x}^i_{k|k-1} + B\,u^{x_i}_{k-1} + K^{x_i}_k\Big[z^{x_i}_k - z^{x_i}_{k|k-1}\Big],\\[1mm]
P^{x_i}_{k|k-1} &= A\,P^{x_i}_{k-1|k-1}A^T + Q^{x_i},\\[1mm]
K^{x_i}_k &= P^{x_i}_{k|k-1}H^T\Big(HP^{x_i}_{k|k-1}H + R^{x_i}\Big)^{-1},\\[1mm]
P^{x_i}_{k|k} &= (I - K^{x_i}_k H)P^{x_i}_{k|k-1},
\end{aligned}
\end{equation}
and similarly for the $y$-axis:
\begin{equation} \label{eq:ekfY}
\begin{aligned}
\hat{y}^i_{k|k} &= A\,\hat{y}^i_{k|k-1} + B\,u^{y_i}_{k-1} + K^{y_i}_k\Big[z^{y_i}_k - z^{y_i}_{k|k-1}\Big],\\[1mm]
P^{y_i}_{k|k-1} &= A\,P^{y_i}_{k-1|k-1}A^T + Q^{y_i},\\[1mm]
K^{y_i}_k &= P^{y_i}_{k|k-1}H^T\Big(HP^{y_i}_{k|k-1}H + R^{y_i}\Big)^{-1},\\[1mm]
P^{y_i}_{k|k} &= (I - K^{y_i}_k H)P^{y_i}_{k|k-1},
\end{aligned}
\end{equation}
where the system matrices are $A=1$ and $B=h$. These filtering equations effectively fuse sensor measurements and attenuate the effect of noise, thus yielding robust state estimates that serve as the basis for the subsequent rendezvous control design.

By leveraging the Kronecker product notation and the network Laplacian $\mathcal{L}$ that captures the inter-robot communication topology, the collective estimation dynamics across all robots can be compactly represented as:
\begin{equation} \label{eq:globalX}
\hat{\bar{x}}_{k+1|k+1} = (I_N \otimes A)\hat{\bar{x}}_{k|k} + (\mathcal{L} \otimes B) u^x_k + \bar{K}^x_{k+1}\Big[\bar{z}^x_{k+1} - \bar{z}^x_{k+1|k}\Big],
\end{equation}
\begin{equation} \label{eq:globalY}
\hat{\bar{y}}_{k+1|k+1} = (I_N \otimes A)\hat{\bar{y}}_{k|k} + (\mathcal{L} \otimes B) u^y_k + \bar{K}^y_{k+1}\Big[\bar{z}^y_{k+1} - \bar{z}^y_{k+1|k}\Big],
\end{equation}
where $\hat{\bar{x}}_{k|k}$, $\hat{\bar{y}}_{k|k}$, $u^x_k$, $u^y_k$, $\bar{K}^x_k$, and $\bar{K}^y_k$ denote the stacked state estimates, control inputs, and Kalman gains for the entire robot team, respectively.

To drive the robots to rendezvous at a desired common point (assumed to be the origin, i.e., $x_0=y_0=0$ for simplicity), we introduce a distributed control law for the $i$th robot:
\begin{equation} \label{eq:controlLaw}
\left\{
\begin{aligned}
u^{x_i}_k &= -L_k\sum_{j=1}^{N}a^{ij}_k\Big(\hat{x}^i_{k|k} - \hat{x}^j_{k|k}\Big),\\[1mm]
u^{y_i}_k &= -L_k\sum_{j=1}^{N}a^{ij}_k\Big(\hat{y}^i_{k|k} - \hat{y}^j_{k|k}\Big),
\end{aligned}
\right.
\end{equation}
where $L_k>0$ is a time-varying gain designed to optimize the overall performance, and $a^{ij}_k$ are the entries of the time-varying adjacency matrix representing the communication links.

The closed-loop dynamics for the state estimates can then be reformulated in a compact form. For example, the $x$-axis update becomes:
\begin{equation} \label{eq:closedLoopX}
\scalebox{0.9}{$
\begin{aligned}
\hat{\bar{x}}_{k+1|k+1} &= (I_N - \bar{K}^x_{k+1}(I_N \otimes H))\Big[(I_N \otimes A)\hat{\bar{x}}_{k|k} + (\mathcal{L} \otimes B) u^x_k\Big] 
 \\  + \bar{K}^x_{k+1}\bar{z}^x_{k+1}
&= (I_N \otimes A)\hat{\bar{x}}_{k|k} + (\mathcal{L} \otimes B) u^x_k + \vartheta^x_k \eta^x_k,
\end{aligned}
$}
\end{equation}
where $\vartheta^x_k$ and $\eta^x_k$ are composite terms that encapsulate the estimation error, process noise, and measurement noise. A similar expression is obtained for the $y$-axis dynamics.

Defining the rendezvous error signals as
\[
\hat{\bar{\xi}}_k = \hat{\bar{x}}_{k|k} - x_0, \quad \hat{\bar{\zeta}}_k = \hat{\bar{y}}_{k|k} - y_0,
\]
with $x_0=y_0=0$, our aim is to design the gain sequences $\{L_k\}$ such that the following quadratic cost functions are minimized:
\begin{equation} \label{eq:costX}
J^x = \mathbb{E}\Bigg[\hat{\bar{\xi}}_M^T Q_M^x\, \hat{\bar{\xi}}_M + \sum_{k=0}^{M-1}\Big(\hat{\bar{\xi}}_k^T \bar{Q}^x\, \hat{\bar{\xi}}_k + {u^x_k}^T \bar{R}^x\, u^x_k\Big)\Bigg],
\end{equation}
\begin{equation} \label{eq:costY}
J^y = \mathbb{E}\Bigg[\hat{\bar{\zeta}}_M^T Q_M^y\, \hat{\bar{\zeta}}_M + \sum_{k=0}^{M-1}\Big(\hat{\bar{\zeta}}_k^T \bar{Q}^y\, \hat{\bar{\zeta}}_k + {u^y_k}^T \bar{R}^y\, u^y_k\Big)\Bigg],
\end{equation}
where $Q_M^x$, $Q_M^y$, $\bar{Q}^x$, $\bar{Q}^y$, $\bar{R}^x$, and $\bar{R}^y$ are appropriately chosen symmetric weighting matrices that penalize state deviations and control energy.
\section{Stochastic LQG Controller Design}
In this section, we derive a distributed stochastic LQG control law that drives the multi-robot system to rendezvous (i.e., converge to a common location) in the presence of stochastic noise. Our derivation builds on the following lemma from \cite{20}.

\begin{lemma} (\cite{20}) \label{lemma1}
Let $x\in \mathbb{R}^n$ be a Gaussian random vector with mean $m$ and covariance $R$, and let $S\in \mathbb{R}^{n\times n}$ be a (possibly stochastic) matrix. Then,
\begin{equation} \label{eq:lemma}
\mathbb{E}\bigl[x^T S x\bigr] = m^T\mathbb{E}(S)m + \mathrm{tr}\bigl(\mathbb{E}(S)R\bigr).
\end{equation}
\end{lemma}

\begin{remark} (\cite{20})
If $S$ is deterministic, the expectation reduces to
\begin{equation} \label{eq:remark}
\mathbb{E}\bigl[x^T S x\bigr] = m^T S m + \mathrm{tr}(S R).
\end{equation}
\end{remark}

Based on the system dynamics in (\ref{eq:discreteDynamics}) and the distributed state estimation discussed earlier, we now formulate the rendezvous problem. Define the error vectors for the $x$ and $y$ coordinates as
\[
\hat{\bar{\xi}}_k = \hat{\bar{x}}_{k|k} - x_0 \quad \text{and} \quad \hat{\bar{\zeta}}_k = \hat{\bar{y}}_{k|k} - y_0,
\]
where the desired rendezvous location is assumed to be $x_0 = y_0 = 0$ without loss of generality. Our objective is to design control inputs that minimize the following quadratic cost functions:
\begin{equation} \label{eq:costX_rev}
J^x = \mathbb{E}\left[\hat{\bar{\xi}}_M^T Q_M^x\, \hat{\bar{\xi}}_M + \sum_{k=0}^{M-1} \left(\hat{\bar{\xi}}_k^T \bar{Q}^x\, \hat{\bar{\xi}}_k + {u^x_k}^T \bar{R}^x\, u^x_k\right)\right],
\end{equation}
\begin{equation} \label{eq:costY_rev}
J^y = \mathbb{E}\left[\hat{\bar{\zeta}}_M^T Q_M^y\, \hat{\bar{\zeta}}_M + \sum_{k=0}^{M-1} \left(\hat{\bar{\zeta}}_k^T \bar{Q}^y\, \hat{\bar{\zeta}}_k + {u^y_k}^T \bar{R}^y\, u^y_k\right)\right],
\end{equation}
with the weighting matrices $Q_M^x$, $\bar{Q}^x$, $\bar{R}^x$ (and similarly for the $y$-axis) chosen to reflect design priorities. 

We now state our main result for the rendezvous control protocol for the $x$-coordinate (a similar result holds for the $y$-coordinate):

\begin{theorem} \label{thm:rendezvous}
For the discrete-time system (\ref{eq:discreteDynamics}) affected by stochastic noise, there exists an optimal distributed rendezvous control law of the form
\begin{equation} \label{eq:controlLaw_rev}
u^x_k = -L_k \hat{\bar{\xi}}_{k|k},
\end{equation}
which minimizes the cost function in (\ref{eq:costX_rev}). The optimal feedback gain $L_k$ is given by
\begin{equation} \label{eq:Lk_rev}
\begin{split}
L_k = \left[\bar{R}^x + \mathbb{E}\Big\{ (\mathcal{L}\otimes B)^T \Pi_{k+1} (\mathcal{L}\otimes B) \Big\}\right]^{-1} \\ \times \mathbb{E}\Big\{ (\mathcal{L}\otimes B)^T \Pi_{k+1} (I_N \otimes A) \Big\},
\end{split}
\end{equation}
with the cost-to-go matrices $\{\Pi_k\}$ computed recursively by
\begin{equation} \label{eq:Ric_rev}
\begin{split}
\Pi_k = \mathbb{E}\Big\{ \Big[(I_N \otimes A) - (\mathcal{L}\otimes B)L_k\Big]^T \Pi_{k+1} \Big[(I_N \otimes A) \\ - (\mathcal{L}\otimes B)L_k\Big] \Big\}  + L_k^T \bar{R}^x L_k + \bar{Q}^x,
\end{split}
\end{equation}
with the terminal condition
\begin{equation} \label{eq:Terminal_rev}
\Pi_M = Q_M^x.
\end{equation}
Furthermore, the minimum value of the cost function is obtained as
\begin{equation} \label{eq:Jmin_rev}
\scalebox{0.9}{$
\begin{split}
J^x_{k} &= \hat{\bar{\xi}}_{M}^T \Pi_M \hat{\bar{\xi}}_{M} + \mathrm{Tr}(\Pi_M \bar{X}_M) \\[1mm]
&\quad + \sum_{k=0}^{M-1} \Big\{ \hat{\bar{\xi}}_{k|k}^T \Big[\mathbb{E}\Big\{ \Big((I_N \otimes A) - (\mathcal{L}\otimes B)L_k\Big)^T \Pi_{k+1} \\
&\quad \times \Big((I_N \otimes A) - (\mathcal{L}\otimes B)L_k\Big) \Big\} + L_k^T \bar{R}^x L_k + \bar{Q}^x\Big] \hat{\bar{\xi}}_{k|k} \\
&\quad + \mathrm{Tr}\Big\{ (I_N \otimes A)^T (I_N \otimes H)^T {K^x_{k+1}}^T \Pi_{k+1} K^x_{k+1}(I_N \otimes H) \bar{P}^{x}_{k|k} \Big\} \\
&\quad + \mathrm{Tr}\Big\{ {K^x_{k+1}}^T \Pi_{k+1} K^x_{k+1} \bar{V}^{x} \Big\} \\
&\quad + \mathrm{Tr}\Big\{ (I_N \otimes H)^T {K^x_{k+1}}^T \Pi_{k+1} K^x_{k+1}(I_N \otimes H) \bar{W}^{x} \Big\} \Big\}.
\end{split}
$}
\end{equation}
\end{theorem}

\begin{proof}
Let $D^x_k = \{\bar{z}^x_1,\ldots,\bar{z}^x_k, u^x_0,\ldots,u^x_{k-1}\}$ denote the data history (sensor measurements and control inputs) available to the robots at time $k$. The optimal cost-to-go function satisfies the Bellman equation
\begin{equation} \label{eq:Bellman_rev}
J_k = \min_{u^x_k} \mathbb{E}\Big\{ \hat{\bar{\xi}}_k^T \bar{Q}^x \hat{\bar{\xi}}_k + {u^x_k}^T \bar{R}^x u^x_k + J_{k+1} \,\Big|\, D^x_k \Big\}.
\end{equation}
Since the filtered state $\hat{\bar{\xi}}_{k|k}$ is a sufficient statistic for the conditional distribution of the true state, we can rewrite (\ref{eq:Bellman_rev}) in terms of $\hat{\bar{\xi}}_{k|k}$:
\begin{equation} \label{eq:Bellman_suff_rev}
J_k = \min_{u^x_k} \mathbb{E}\Big\{ \hat{\bar{\xi}}_k^T \bar{Q}^x \hat{\bar{\xi}}_k + {u^x_k}^T \bar{R}^x u^x_k + J_{k+1} \,\Big|\, \hat{\bar{\xi}}_{k|k} \Big\}.
\end{equation}
At the final time step, the cost is given by
\begin{equation} \label{eq:TerminalCost_rev}
J_M = \mathbb{E}\Big\{ \hat{\bar{\xi}}_M^T Q_M^x \hat{\bar{\xi}}_M \,\Big|\, \hat{\bar{\xi}}_{M|M} \Big\}.
\end{equation}
Using Lemma~\ref{lemma1}, we express this cost in terms of the filtered state and its covariance. By induction, one can show that the optimal cost function admits a quadratic form:
\begin{equation} \label{eq:QuadraticCost_rev}
J_k = \mathbb{E}\Big\{ \hat{\bar{\xi}}_{k|k}^T \Pi_k \hat{\bar{\xi}}_{k|k} \Big\} + \Psi_k,
\end{equation}
where $\Pi_k$ and $\Psi_k$ are determined recursively. Following standard dynamic programming techniques and applying Lemma~\ref{lemma1} to handle the stochastic terms (which include the process and measurement noise contributions), one arrives at the recursive equation in (\ref{eq:Ric_rev}) with terminal condition (\ref{eq:Terminal_rev}). The resulting optimal control input is then given by (\ref{eq:controlLaw_rev}), which, upon substitution, yields the minimum cost as in (\ref{eq:Jmin_rev}). The detailed derivation involves standard completion-of-squares arguments and can be found in related works (e.g., \cite{21}), and is omitted here for brevity.
\end{proof}
\section{Stability Analysis}
To study the stability of the discrete-time system (\ref{eq:discreteDynamics}) under the optimal rendezvous protocol, we begin with a useful result.

\begin{lemma} (\cite{19}) \label{lem:stability}
Assume there exists a stochastic process $V_k(\zeta_k)$ and positive constants $\underline{\kappa}$, $\bar{\kappa}$, $\mu > 0$, and $0 < \sigma \leq 1$ such that
\begin{equation} \label{eq:boundV}
\underline{\kappa} \|\zeta_k\|^2 \leq V_k(\zeta_k) \leq \bar{\kappa} \|\zeta_k\|^2,
\end{equation}
and
\begin{equation} \label{eq:decreaseV}
\mathbb{E}\{V_k(\zeta_k) \mid \zeta_{k-1}\} - V_{k-1}(\zeta_{k-1}) \leq \mu - \sigma V_{k-1}(\zeta_{k-1}).
\end{equation}
Then, the process $\{\zeta_k\}$ is exponentially bounded in the mean square sense, i.e.,
\begin{equation} \label{eq:expBound}
\mathbb{E}\{\|\zeta_k\|^2\} \leq \frac{\bar{\kappa}}{\underline{\kappa}} \, \mathbb{E}\{\|\zeta_0\|^2\}(1-\sigma)^k + \frac{\mu}{\underline{\kappa}} \sum_{i=1}^{k-1} (1-\sigma)^i,
\end{equation}
and the process is bounded almost surely.
\end{lemma}

We now analyze the stability of the discrete-time system under the proposed distributed rendezvous protocol. Let the rendezvous error for the $x$-coordinate be defined as
\[
\bar{\xi}_{k|k} = \bar{x}_{k|k} - x_0,
\]
where $x_0$ is the desired rendezvous location (assumed to be zero without loss of generality). The following theorem establishes the exponential mean square boundedness of the rendezvous error.

\begin{theorem} \label{thm:stability}
Consider the discrete-time system (\ref{eq:discreteDynamics}) with the distributed rendezvous protocol given by
\begin{equation} \label{eq:control_rendezvous}
u^x_k = -L_k \hat{\bar{\xi}}_{k|k}.
\end{equation}
Suppose that the error covariance satisfies
\begin{equation} \label{eq:covarianceBound}
P^x_{k+1|k+1} \leq P^x_{k+1|k} \leq \bar{p},
\end{equation}
and there exists $\varepsilon > 0$ such that
\[
\mathbb{E}\{\|\bar{\xi}_{0|0}\|^2\} \leq \varepsilon.
\]
Then, the rendezvous error is exponentially bounded in the mean square sense, i.e., for all $k \geq 0$,
\begin{equation} \label{eq:rendezvousBound}
\mathbb{E}\{\|\bar{\xi}_{k|k}\|^2\} \leq \frac{\bar{\kappa}}{\underline{\kappa}}\, \varepsilon \,(1-\lambda)^{k} + \frac{\mu}{\underline{\kappa}} \sum_{m=1}^{k-1} (1-\lambda)^{m},
\end{equation}
where $\lambda$ is a positive constant satisfying $0<\lambda \leq 1$, and $\underline{\kappa}$, $\bar{\kappa}$, and $\mu$ are as defined in Lemma~\ref{lem:stability}.
\end{theorem}

\begin{proof}
Define the Lyapunov function candidate for the $x$-coordinate rendezvous error as
\begin{equation} \label{eq:lyapunov}
V_k = \hat{\bar{\xi}}_{k|k}^T \Pi_k \hat{\bar{\xi}}_{k|k},
\end{equation}
where $\Pi_k$ is the positive definite matrix obtained from the Riccati recursion in (\ref{eq:Ric_rev}). Since $\Pi_k > 0$, $V_k$ is positive definite when $\hat{\bar{\xi}}_{k|k} \neq 0$.

Next, we consider the difference of the Lyapunov function along the trajectories:
\begin{equation} \label{eq:lyapDiff}
\begin{split}
\Delta V_k &= \mathbb{E}\{V_{k+1} \mid \hat{\bar{\xi}}_{k|k}, \ldots, \hat{\bar{\xi}}_{0|0}\} - V_k \\
&= \mathbb{E}\{\hat{\bar{\xi}}_{k+1|k+1}^T \Pi_{k+1} \hat{\bar{\xi}}_{k+1|k+1} \mid \hat{\bar{\xi}}_{k|k}, \ldots, \hat{\bar{\xi}}_{0|0}\} - \hat{\bar{\xi}}_{k|k}^T \Pi_k \hat{\bar{\xi}}_{k|k}.
\end{split}
\end{equation}

By substituting the closed-loop dynamics (see, e.g., (\ref{eq:closedLoopX})) into the expression for $\hat{\bar{\xi}}_{k+1|k+1}$, and after applying the expectation operator and using properties of the trace along with Lemma~\ref{lemma1}, we can express the difference as
\begin{equation} \label{eq:lyapDiff2}
\Delta V_k \leq \mu_k - \lambda V_k,
\end{equation}
where $\mu_k$ collects the terms due to process and measurement noise:
\begin{equation} \label{eq:mu_k}
\begin{split}
\mu_k = \, &\mathrm{Tr}\Big\{ (I_N \otimes H)^T {K^x_{k+1}}^T \Pi_{k+1} K^x_{k+1}(I_N \otimes H) \bar{P}^x_{k|k} \Big\} \\
&+ \mathrm{Tr}\Big\{ (I_N \otimes H)^T {K^x_{k+1}}^T \Pi_{k+1} K^x_{k+1}(I_N \otimes H) \bar{W}^x \Big\} \\
&+ \mathrm{Tr}\Big\{ {K^x_{k+1}}^T \Pi_{k+1} K^x_{k+1} \bar{V}^x \Big\}.
\end{split}
\end{equation}

Since $\lambda V_k \leq V_k$, the inequality in (\ref{eq:lyapDiff2}) fits the structure of (\ref{eq:decreaseV}) in Lemma~\ref{lem:stability}. Thus, applying Lemma~\ref{lem:stability} directly, we obtain that
\[
\mathbb{E}\{\|\hat{\bar{\xi}}_{k|k}\|^2\} \leq \frac{\bar{\kappa}}{\underline{\kappa}}\, \mathbb{E}\{\|\hat{\bar{\xi}}_{0|0}\|^2\}(1-\lambda)^{k} + \frac{\mu}{\underline{\kappa}} \sum_{m=1}^{k-1} (1-\lambda)^m.
\]
Noting that the total error includes both the estimation error and the filter covariance, i.e.,
\begin{equation} \label{eq:totalError}
\mathbb{E}\{\|\bar{\xi}_{k|k}\|^2\} = \mathbb{E}\{\|\hat{\bar{\xi}}_{k|k}\|^2\} + \mathrm{Tr}(P^x_{k|k}),
\end{equation}
and assuming that the initial estimation error is bounded, the overall rendezvous error is exponentially bounded in the mean square sense. This completes the proof.
\end{proof}

\begin{remark}
From Theorem~\ref{thm:stability}, it is evident that the upper bound of $\mu_k$ is directly influenced by the covariances of the process and measurement noise. These noise characteristics, in turn, determine the upper bound of the mean square rendezvous error $\mathbb{E}\{\|\bar{\xi}_{k|k}\|^2\}$.
\end{remark}
\begin{algorithm}
\caption{Distributed LQG Rendezvous Control for a Group of Mobile Robots With First-Order Dynamics}
\label{alg:rendezvous}
\begin{algorithmic}[1]
    \State \textbf{Input:} For each robot $i\in\{1,\ldots,N\}$ and its neighbors $j\in\vartheta\setminus\{i\}$, initialize:
    \begin{itemize}
        \item True positions: $\mathbf{x}_i(0)\in\mathbb{R}^{n_i}$, $\mathbf{y}_i(0)\in\mathbb{R}^{n_i}$;
        \item Initial state estimates: $\hat{\mathbf{x}}^i_{0|0}\in\mathbb{R}^{n_i}$, $\hat{\mathbf{y}}^i_{0|0}\in\mathbb{R}^{n_i}$;
        \item Error covariance matrices: $\mathbf{P}^{x_i}_{0|0}\in\mathbb{S}^{n_i}$, $\mathbf{P}^{y_i}_{0|0}\in\mathbb{S}^{n_i}$;
        \item A small tolerance parameter $\epsilon\ll 0.05$.
    \end{itemize}
    
    \State \textbf{Output:} Control inputs (velocities) $\mathbf{v}^{x}_i(k)$, $\mathbf{v}^{y}_i(k)$ and updated positions $\mathbf{x}_i(k)$, $\mathbf{y}_i(k)$.
    
    \While {$(\|\mathbf{x}_i(k)-\mathbf{x}_j(k)\| > \epsilon)$ \textbf{or} $(\|\mathbf{y}_i(k)-\mathbf{y}_j(k)\| > \epsilon)$}
    \State \textbf{State Estimation:} Using onboard sensors (e.g., odometry, IMU), each robot $i$ updates its state estimates via the Kalman filter:
    \begin{align*}
        \hat{\mathbf{x}}^i_{k|k} &= \mathbf{A}\hat{\mathbf{x}}^i_{k|k-1} + \mathbf{B} \, \mathbf{u}^{x_i}_{k-1} + \mathbf{K}^{x_i}_{k}\Big[\mathbf{z}^{x_i}_k - \mathbf{C}\hat{\mathbf{x}}^i_{k|k-1}\Big], \\
        \mathbf{P}^{x_i}_{k|k-1} &= \mathbf{A}\mathbf{P}^{x_i}_{k-1|k-1}\mathbf{A}^T + \mathbf{Q}^{x_i},\\[1mm]
        \mathbf{K}^{x_i}_{k} &= \mathbf{P}^{x_i}_{k|k-1}\mathbf{C}^T\Big[\mathbf{C}\mathbf{P}^{x_i}_{k|k-1}\mathbf{C}^T + \mathbf{R}^{x_i}\Big]^{-1},\\[1mm]
        \mathbf{P}^{x_i}_{k|k} &= \bigl(\mathbf{I}-\mathbf{K}^{x_i}_{k}\mathbf{C}\bigr)\mathbf{P}^{x_i}_{k|k-1},
    \end{align*}
    and similarly for the $y$-coordinate.
    
    \State \textbf{Local Riccati Update:} For each coordinate, compute the cost-to-go matrix by solving the following backward Riccati equation:
    \begin{align*}
        \Pi^{x}_{k} &= \bigl(\mathbf{A} - \mathbf{B}L^{x}_k\bigr)^T \Pi^{x}_{k+1}\bigl(\mathbf{A} - \mathbf{B}L^{x}_k\bigr) + L^{x^T}_k \mathbf{R}^{x}L^{x}_k + \mathbf{Q}^{x},\\[1mm]
        \Pi^{y}_{k} &= \bigl(\mathbf{A} - \mathbf{B}L^{y}_k\bigr)^T \Pi^{y}_{k+1}\bigl(\mathbf{A} - \mathbf{B}L^{y}_k\bigr) + L^{y^T}_k \mathbf{R}^{y}L^{y}_k + \mathbf{Q}^{y},
    \end{align*}
    with terminal conditions $\Pi^{x}_{M} = Q_M^x$ and $\Pi^{y}_{M} = Q_M^y$.
    
    \State \textbf{Compute Optimal Gains:} Determine the control gains for the $x$- and $y$-directions:
    \begin{align*}
        L^{x}_k &= \Bigl[\mathbf{R}^{x} + \mathbf{B}^T \Pi^{x}_{k+1} \mathbf{B}\Bigr]^{-1}\mathbf{B}^T \Pi^{x}_{k+1} \mathbf{A},\\[1mm]
        L^{y}_k &= \Bigl[\mathbf{R}^{y} + \mathbf{B}^T \Pi^{y}_{k+1} \mathbf{B}\Bigr]^{-1}\mathbf{B}^T \Pi^{y}_{k+1} \mathbf{A}.
    \end{align*}
    
    \State \textbf{Communication:} Robot $i$ exchanges its estimated states $\hat{\mathbf{x}}^{i}_{k|k}$ and $\hat{\mathbf{y}}^{i}_{k|k}$ with neighboring robots $j\in\vartheta\setminus\{i\}$.
    
    \State \textbf{Rendezvous Update:} Update the robot positions using the distributed LQG rendezvous protocol:
    \begin{align*}
        \mathbf{x}_i(k+1) &= \mathbf{x}_i(k) - L^{x}_k \sum_{j=1}^{N} a_{ij}(k) \Bigl(\hat{\mathbf{x}}^{i}_{k|k} - \hat{\mathbf{x}}^{j}_{k|k}\Bigr),\\[1mm]
        \mathbf{y}_i(k+1) &= \mathbf{y}_i(k) - L^{y}_k \sum_{j=1}^{N} a_{ij}(k) \Bigl(\hat{\mathbf{y}}^{i}_{k|k} - \hat{\mathbf{y}}^{j}_{k|k}\Bigr),
    \end{align*}
    where $a_{ij}(k)$ are the entries of the communication adjacency matrix at time $k$.
    
    \State \textbf{Compute Velocities:} Derive the corresponding speed commands:
    \begin{align*}
        \mathbf{v}^{x}_i(k) &= \frac{\mathbf{x}_i(k+1)-\mathbf{x}_i(k)}{h}, \\
        \mathbf{v}^{y}_i(k) &= \frac{\mathbf{y}_i(k+1)-\mathbf{y}_i(k)}{h}.
    \end{align*}
    
    \State Update time index: $k \leftarrow k+1$.
    \EndWhile
\end{algorithmic}
\end{algorithm}

\section{Experimental Validation}

\begin{figure}[h!] 
  \begin{center}
\includegraphics[width=0.4 \textwidth]{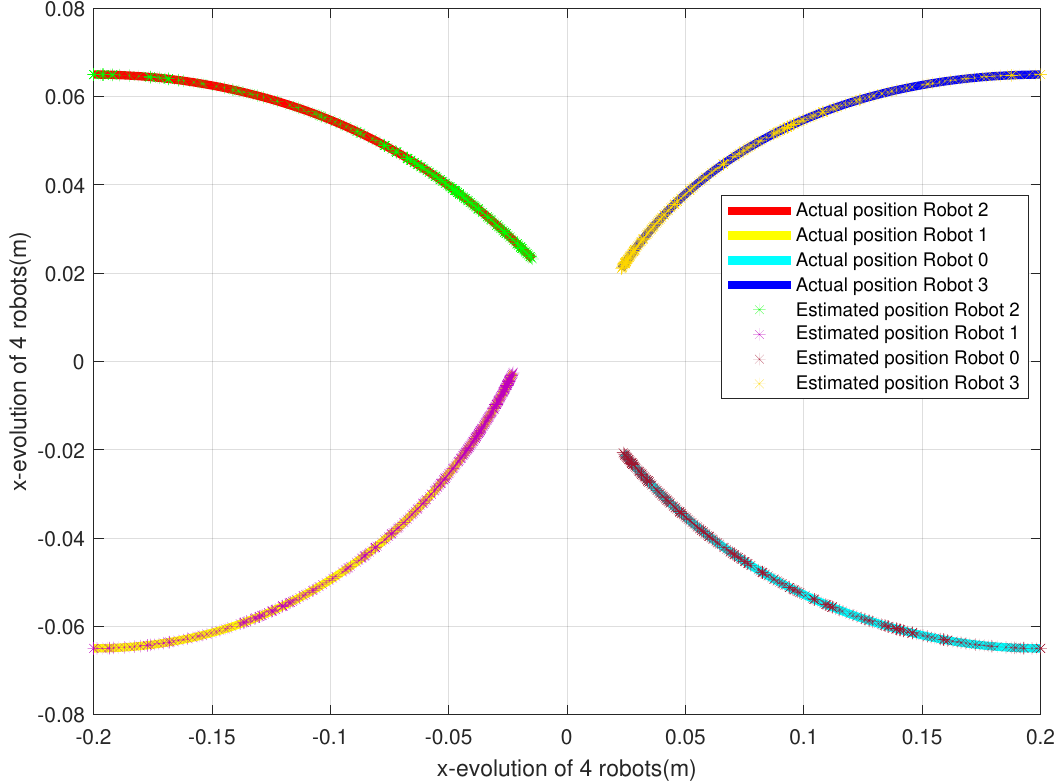}
  \end{center}
  \caption{Experimental testing results of the proposed LQG consensus control: x-evolution and y-evolution of four mobile robots considering small noisy sensor measurements } 
  \label{Fig3}
\end{figure} 

\begin{figure}[h!] 
  \begin{center}
\includegraphics[width=0.4 \textwidth]{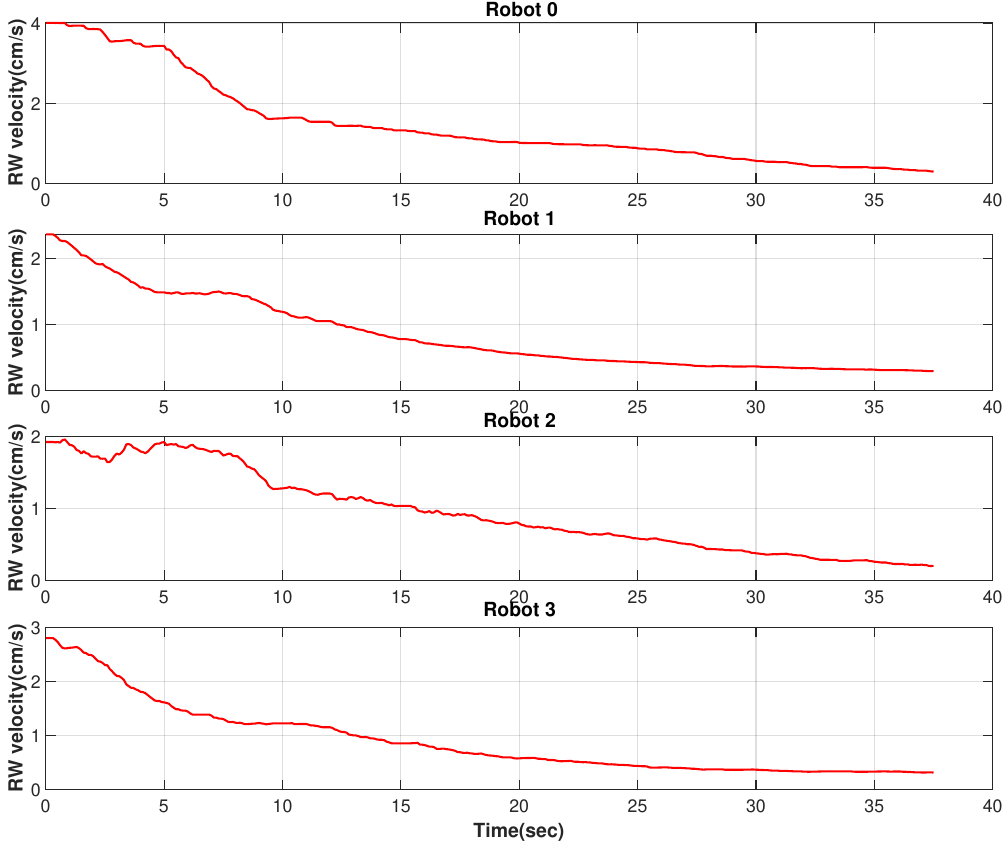}
  \end{center}
  \caption{Experimental testing results of the proposed LQG consensus control: right wheel velocities of four mobile robots considering small noisy sensor measurements } 
  \label{Fig3}
\end{figure} 

\begin{figure}[h!] 
  \begin{center}
\includegraphics[width=0.4 \textwidth]{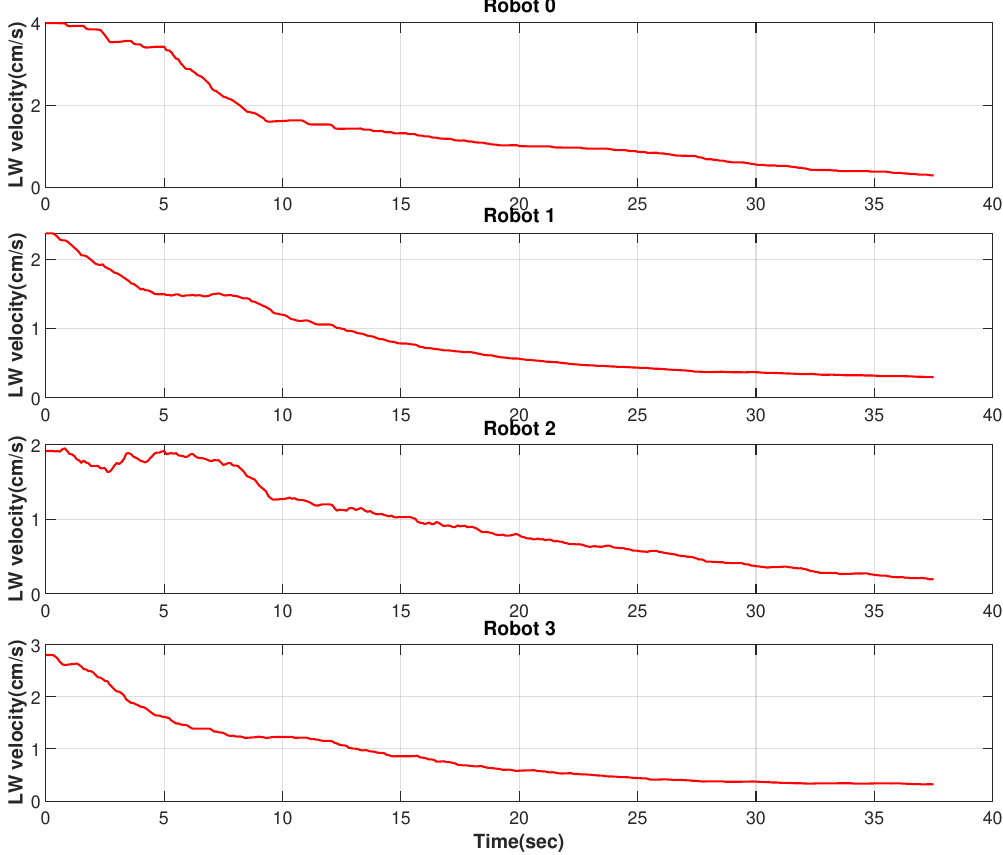}
  \end{center}
  \caption{Experimental testing results of the proposed LQG consensus control: left wheel velocities of four mobile robots considering small noisy sensor measurements } 
  \label{Fig3}
\end{figure} 

\begin{figure}[h!] 
  \begin{center}
\includegraphics[width=0.4 \textwidth]{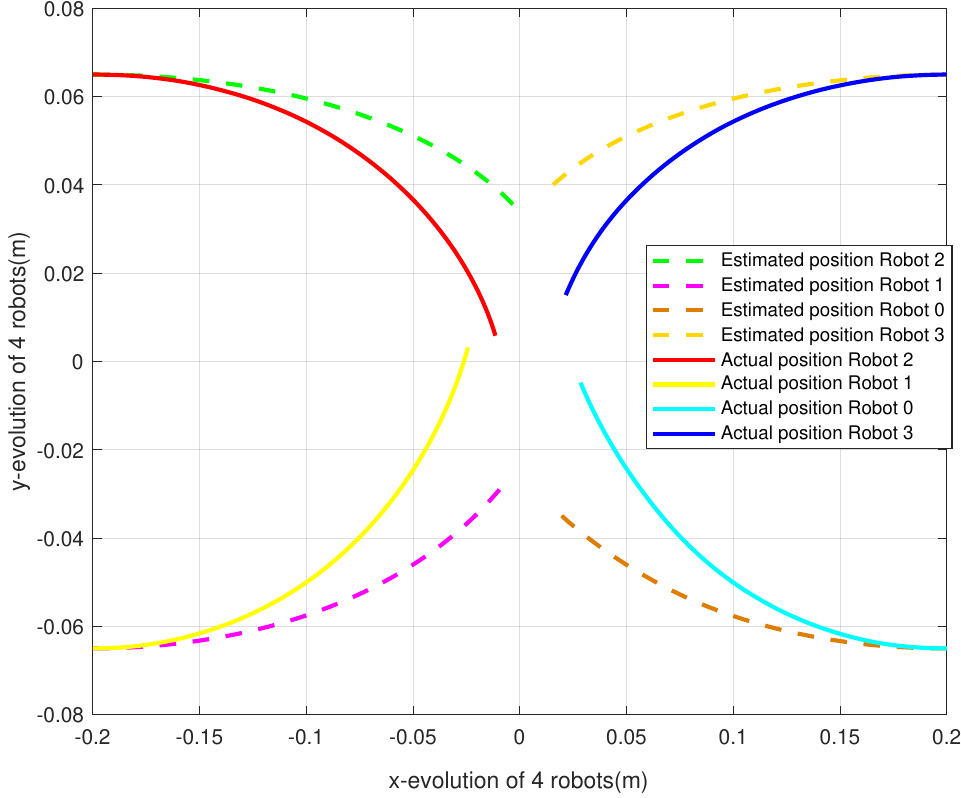}
  \end{center}
  \caption{Experimental testing results of the proposed LQG consensus control: x-evolution and y-evolution of four mobile robots considering high noisy sensor measurements} 
  \label{Fig3}
\end{figure} 

\begin{figure}[h!] 
  \begin{center}
\includegraphics[width=0.4 \textwidth]{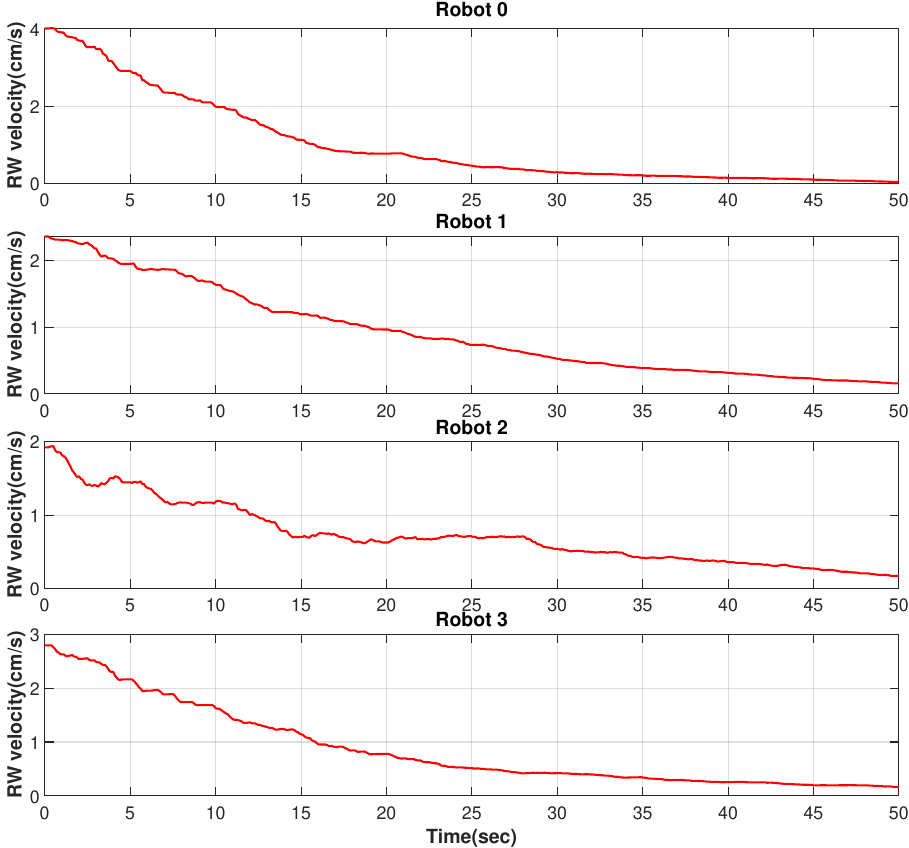}
  \end{center}
  \caption{Experimental testing results of the proposed LQG consensus control: right wheel velocities of four mobile robots considering high noisy sensor measurements} 
  \label{Fig3}
\end{figure} 

\begin{figure}[h!] 
  \begin{center}
\includegraphics[width=0.5 \textwidth]{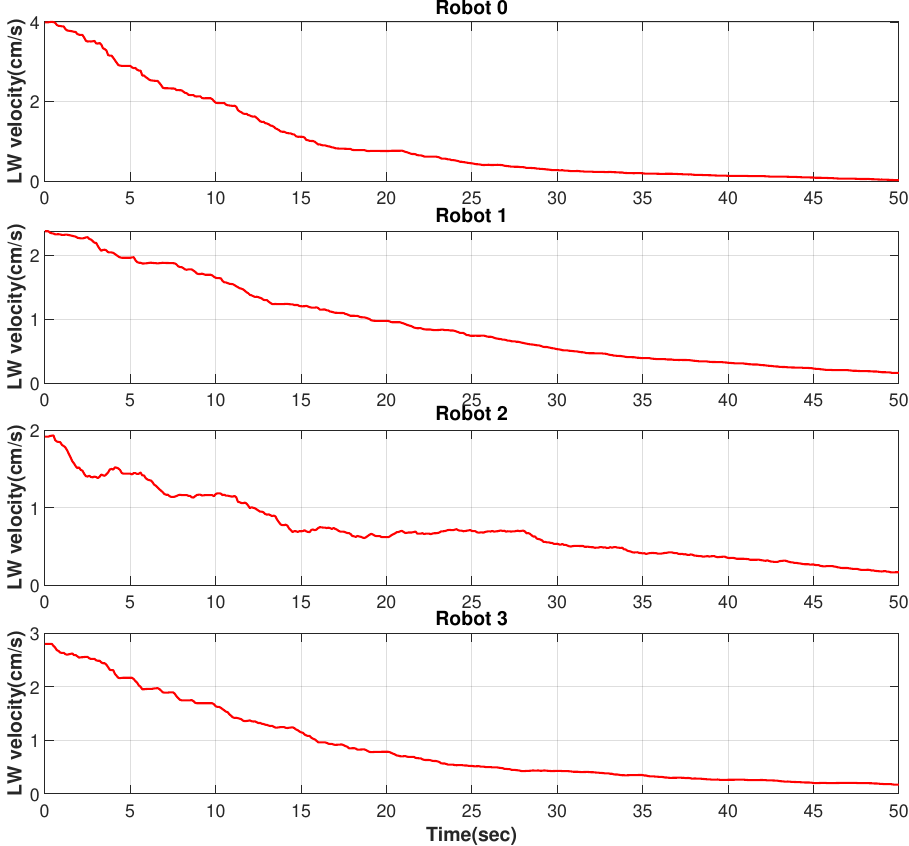}
  \end{center}
  \caption{Experimental testing results of the proposed LQG consensus control: left wheel velocities of four mobile robots considering high noisy sensor measurements} 
  \label{Fig3}
\end{figure} 
In this section, we experimentally validate the effectiveness of the proposed distributed LQG optimal control strategy for achieving rendezvous among mobile robots using the Robotrium platform. Each robot employs an extended Kalman filter (EKF) to fuse measurements from multiple sources—specifically, wheel odometry and inertial measurement unit (IMU) sensors—to estimate its pose accurately. Since these sensors operate with different latencies, the EKF processes buffered measurements to update the pose estimates in real time. Inter-robot communication is facilitated via a Bluetooth network, which ensures timely data exchange among agents.

The robots are initially placed at random positions with zero initial velocities. For instance, the initial positions are set as follows: 
\[
\begin{aligned}
x_{0}(0) &= 0.2,\quad y_{0}(0) = -0.065,\\
x_{1}(0) &= -0.2,\quad y_{1}(0) = -0.065,\\
x_{2}(0) &= -0.2,\quad y_{2}(0) = 0.065,\\
x_{3}(0) &= 0.2,\quad y_{3}(0) = 0.065.
\end{aligned}
\]
The initial estimated poses are set equal to the actual initial poses. Additionally, the wheel speeds are constrained such that $\|v_l\|\leq15.4$~cm/s and $\|v_r\|\leq15.4$~cm/s.

 Figure~5 presents snapshots of the robot positions over time. As observed, the robots successfully converge to the agreed rendezvous point, computed as the average of their initial positions, i.e., 
\[
x_{\mathrm{rendezvous}} = \frac{1}{4} \sum_{i=0}^{3} x_i(0) \quad \text{and} \quad y_{\mathrm{rendezvous}} = \frac{1}{4} \sum_{i=0}^{3} y_i(0).
\]

Figures~6, 7, and 8 display the experimental results under two sensor noise scenarios. In the first scenario, with odometry pose covariance $Q_{\mathrm{odom}}=1\times10^{-6}$ and IMU pose covariance $Q_{\mathrm{IMU}}=1\times10^{-4}$, the estimated poses closely track the actual poses, and the robots converge to the rendezvous point. In the second scenario, the noise levels are increased to $Q_{\mathrm{odom}}=1\times10^{-3}$ and $Q_{\mathrm{IMU}}=1\times10^{-2}$. As expected, higher sensor covariance degrades localization performance, leading to increased localization and convergence errors; however, these errors remain bounded. 

Overall, the experimental results on the Robotrium platform confirm that the proposed distributed LQG rendezvous control strategy is robust to sensor noise and effectively drives the multi-robot system to achieve the desired rendezvous.

\section{Conclusion}
This paper presented a distributed LQG optimal control strategy for achieving rendezvous in multi-robot systems. The proposed approach integrates local sensor fusion using an extended Kalman filter and inter-agent communication over a Bluetooth network to enhance localization and coordination. We experimentally validated the effectiveness of the control strategy using the Robotrium platform, demonstrating its capability to drive robots to a common rendezvous point despite sensor noise and communication delays.

The experimental results confirmed that the estimated poses closely track actual robot positions and that the proposed method ensures convergence to the rendezvous point. Furthermore, the study investigated the impact of sensor noise on localization accuracy and convergence performance. While increased sensor covariance led to degraded localization accuracy, the overall system remained stable, with bounded localization and rendezvous errors.

Future research directions include extending the approach to dynamic rendezvous scenarios where the target location evolves over time, investigating the impact of network latency on system performance, and incorporating more sophisticated communication protocols to enhance robustness in real-world deployments.

\end{document}